# A Comparative Study of Audio Compression Based on Compressed Sensing and Sparse Fast Fourier Transform (SFFT): Performance and Challenges

Hossam M.Kasem, Maha El-Sabrouty
Electronic and Communication Engineering, Egypt-Japan University of Science and Technology (EJUST)
New Borg El-Arab, Egypt
hossam.kasem@ejust.edu.eg, maha.elsabrouty@ejust.edu.eg

*Abstract*—Audio compression has become one of the basic multimedia technologies. Choosing an efficient compression scheme that is capable of preserving the signal quality while providing a high compression ratio is desirable in the different standards worldwide. In this paper we study the application of two highly acclaimed sparse signal processing algorithms, namely, Compressed Sensing (CS) and Sparse Fart Fourier transform, to audio compression. In addition, we present a Sparse Fast Fourier transform (SFFT)-based framework to compress audio signal. This scheme embeds the K-largest frequencies indices as part of the transmitted signal and thus saves in the bandwidth required for transmission.

## I. INTRODUCTION

Audio compression has witnessed great progress thanks to several revolutionary research discoveries. Paying attention to the audio signal nature has contributed to make the shift from the basic sampling, quantization and coding schemes to the more efficient compression of Linear Prediction Coding (LPC) and Code Excited Linear prediction coding (CELP) and other model based systems for speech signal which depends on the vocal tract model [1]. The perceptual audio compression comes as another paradigm shift in the audio technology, where high quality low bit rate coding is achieved through carefully studying the human auditory model to achieve high compression ratio without discomforting the listening experience[2],[3]. Recently, sparseness of audio signal has been exploited with the aim of achieving even higher compression ratio than the current compression techniques used in the multimedia coding standards [4].

Sparse signal processing is an active research area that has recently witnessed two major research trends, namely compressed sampling (sensing) and Sparse Fast Fourier Transform. Compressed sensing pioneered by the work of Cand`es and Donoho [5],[6] aims at providing a universal framework for randomly sampling and representing sparse signals using linear measurements at a sampling rate much smaller than the Nyquist rate. The main motivation behind the work is that random sampling will introduce incoherent noise [7] and by the knowledge of the measurement matrix used to produce the random samples, the original signal can be recovered at the decoder side. Compressed sensing has found its way to many applications in imaging, and more recently in audio compression [8], [9].

On the other hand, Sparse Fast Fourier Transform is more related to the signal content as opposed to the universal random sampling compressed sensing. Sparse FFT searches the sparse FFT domain for the pre-determined number largest coefficients in magnitude. The most appealing feature of the Sparse Fast Fourier Transform (SFFT) is that it is capable of estimating the largest coefficients of the Fourier transform of complex vector x of length $N$ in sub-linear time $O(\log N \sqrt{Nk \log N})$ [10].

In this paper we compare the two sparse processing systems: The universal compressed sensing and the signal intrinsic-based SFFT. We summarize the challenges and the possible areas of utilization of each system for the case study of audio coding. In fact, we cast the Sparse FFT family of algorithms as a special case of compressed sensing. In addition, we try to quantify the performance gain achieved by considering the signal content in the SFFT compared to the general compressed sensing system through simulation. The gain noticed in the simulations indicates the worth of investing effort in designing new class of sensing matrices custom to the data types to take into account the proper signal model in addition to the sparsity property.

This paper is organized as follows section II gives a brief overview of CS, its Reconstruction algorithms, its advantages and limitations. Section III presents an overview of SFFT algorithm and manipulating it as a special case of compressed sensing. Section IV presents our audio compression methodology. The experimental results are presented in section V and the conclusion of this paper is presented in section VI.

Notation: Matrices and vectors are typefaced using slanted bold uppercase and lowercase letters. $l_0$ norm ($\|.\|_0$) is defined as the number of nonzero components in its argument.

## II. COMPRESSED SENSING

### A. Theory

Suppose we have a vector $\mathbf{x} \in R^N$, which can be represented in a certain domain by transform matrix $\mathbf{\Psi}$ (this transform matrix can be generally viewed as the transform domain transform, i.e. Wavelet Transform (WT),

Discrete Cosine Transform (DCT) and Discrete Fourier Transform (DFT)). The input signal **x** can be represented as:

$$\mathbf{x} = \mathbf{\Psi f} \quad (1)$$

where $\mathbf{\Psi}$ is $N \times N$ matrix whose columns are orthonormal basis functions, and **f** is the coefficients vector. If there are only $K$ non-zero coefficients in $\mathbf{\Psi}$ domain, we can say that **x** is $K$-sparse. According to the CS theory[5],[6], signal **f** can be obtained through a number M of random measurements which can be obtained by random linear projections of **x** over measurement matrix according to the next equation.

$$\mathbf{y}_{CS} = \mathbf{\Phi x} \quad (2)$$

where $\mathbf{y}_{CS}$ is an M-dimensional measurement vector with $M \ll N$ and $\mathbf{\Phi}$ is $M \times N$ random matrix which represents the measurement process. typically $M \geq const \times K \log(\frac{N}{K})$. Substituting Eq.(1) into Eq.(2) we get:

$$\mathbf{y}_{CS} = \mathbf{\Phi \Psi f} = \mathbf{Af} \quad (3)$$

where the sensing matrix **A** is typically full rank. For compressed sensing $\mathbf{\Phi}$ and $\mathbf{\Psi}$ must exhibit the minimal coherence. In order to minimize coherence between the matrix pair $\langle \mathbf{\Psi, \Phi} \rangle$, a sensing (measurement matrix $\mathbf{\Phi}$) is generally constructed random with random orthobases selected from independent and uniformly sampled unit random sphere [11].

To obtain exact recovery, the rule of thumb is to apply incoherent sampling and taking measurements 4 times the sparsity level of the signal [11]. In this case, using appropriate reconstruction algorithm the signal can be exact recovered. This fact is known as four-to-one practical rule.

*B. Reconstraction*

There is a variety of algorithms in literature to reconstruct the sparse signal from its compressed sampled set.

The reconstruction problem can be cast as follows: Given the compressed sampled signal $\mathbf{y}_{CS}$ and the sensing matrix **A** estimate the sparse signal **f** within the class of interest such that $\mathbf{y}_{CS} = \mathbf{Af}$ exactly or approximately. Reconstruction of compressed sensing signal relies on the sparsity property of the signal **f**.

The first set of algorithms is based on the $l_0$ norm, i.e. $\|.\|_0$ and its variants as a measure of sparsity. The $l_0$ norm of a vector is defined as the number of nonzero entries in this vector norm-zero defined. The recovery algorithm can be directly represented as $l_0$ norm minimization problem:

$$\hat{\mathbf{f}} = \mathbf{argmin_f} \|\mathbf{f}\|_0$$
Subject to
$$\mathbf{y}_{CS} = \mathbf{Af} \quad (4)$$

Solving (4) relies on an exhaustive search. The objective function $\|.\|_0$ is no convex and hence (4) is very difficult to solve. An alternative to the $l_0$ norm used in (4) is to use the $l_1$ norm defined as $\|\mathbf{f}\|_1 = \sum_{n=1}^{N} \mathbf{f(n)}$. The resulting adaptation of (4), known as basis pursuit (BP) [12] is formally defined as:

$$\hat{\mathbf{f}} = \arg\min_\mathbf{f} \|\mathbf{f}\|_1$$
Subject to
$$\mathbf{y}_{CS} = \mathbf{Af} \quad (5)$$

Since the $l_1$ norm is convex, (5) can be seen as a convex relaxation of (4). Thanks to the convexity, this algorithm can be implemented as a linear program, making its computational complexity polynomial in the signal length. To solve problem (5) we used CVX, a package for specifying and solving convex programs [13],[14].

Another family of algorithms is the greedy algorithms, such that Orthogonal Matching Pursuit (OMP)[15], stagewise Orthogonal Matching Pursuit (StOMP)[16], CoSaMp[17]. The third approach is combinatorial algorithms which combined both chaining pursuit and HHS (Heavy Hitters on Steroids) pursuit [18],[19].

In This paper we have employed both reconstruction algorithm Greedy algorithm (OMP) and convex relaxation (CVX). The Orthogonal Matching Pursuit (OMP) can be explained in the next flow chart as in Fig.1 [15]. Where $t$ the number of iterations, $K$ is the sparsity level of the signal, **A** is sensing matrix and $N$, $M$ is the length of signal and measurements respectively.

The major advantages Orthogonal Matching Pursuit (OMP) are its ease of implementation and its speed. The running time of the OMP algorithm is dominated by Step 2, whose total cost is $O(KNM)$ at iteration $t$, the least-squares problem can be solved with marginal cost $O(tM)$. To do so, we maintain a QR factorization of $\mathbf{A_t}$.

III. SPARSE FAST FOURIER TRANSFORM (SFFT)

The Fast Fourier Transform (FFT) is one of the most successful transforms, mainly thanks to its efficient structure. FFT, basically, computes the Discrete Fourier Transform (DFT) of an $N$-dimensional signal in $O(N \log N)$ time. The FFT algorithm plays a central role in several application areas, including signal processing and audio/image/video compression. It is also a fundamental subroutine in integer multiplication and encoding/decoding of error-correcting codes [10].

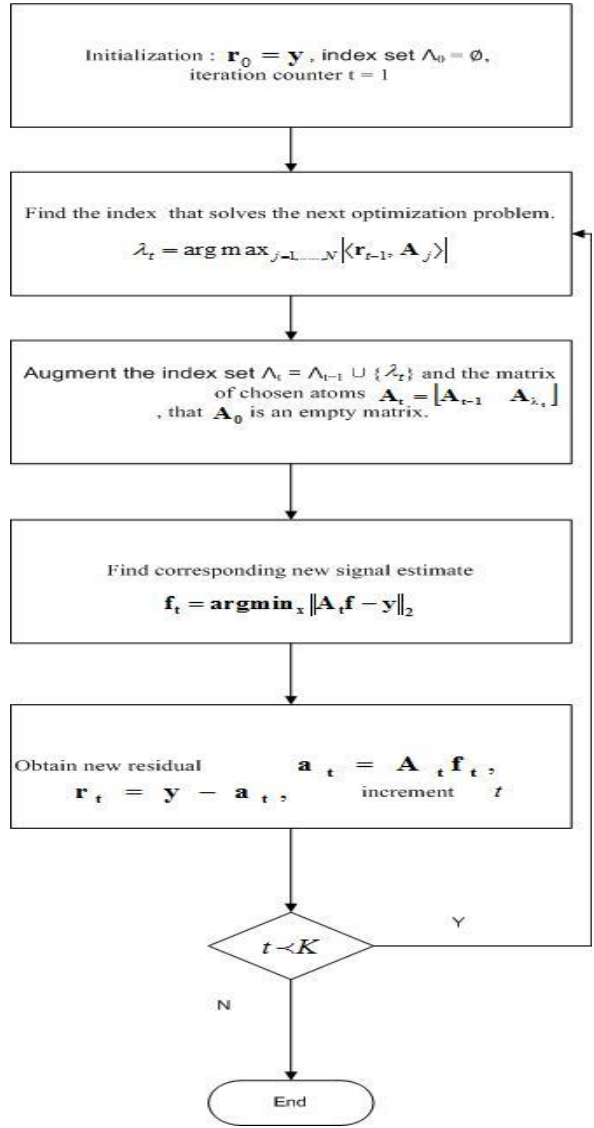

Figure 1. FLOW Chart for OMP

SFFT is a sub-linear algorithm for searching sparse signals in Fourier transform domain for the largest coefficient in magnitude. The main advantage of this algorithm is its simplicity. The algorithm has a simple structure, which leads to efficient runtime. Typically case of $N$ a power of 2 is $O(\log N \sqrt{Nk \log N})$.

The idea behind the algorithm is binning the Fourier coefficients into a small number of buckets. Since the signal is sparse in the frequency domain, each bucket is likely to have only one large coefficient, which can then be located (to find its position) and estimated (to find its value). For the algorithm to be sub-linear, the binning has to be done in sub-linear time. To achieve this goal, these algorithms bin the Fourier coefficient using an $N$-dimensional filter vector (Gaussian function with Dolph- chybyshev function) G that is concentrated both in time and frequency [10]. Once a large coefficients is isolated in a bucket, one needs to identify its frequency. In contrast to past work which typically uses binary search for this task, Specifically, we simply select the set of "large" bins which are likely to contain large coefficients, and directly estimate all frequencies in those bins.

In fact, by choosing only the largest K coefficients in the frequency domain, the sparse FFT can be represent as a special case of compressed sensing where the sparse K vector long signal.

$$\mathbf{X}_s = \mathbf{A}\,\mathbf{X}$$

$$= \begin{bmatrix} 0 & \cdots & 1 & 0 & \cdots & 0 \\ 0 & \cdots & \cdots & 1 & 0 & \cdots \\ \vdots & 0 & \cdots & \cdots & 1 & \cdots \\ \vdots & & 0 & & & 1 \\ & & & \cdots & \cdots & \end{bmatrix} \mathbf{X}$$

The matrix A has one in each row corresponding to the index of the non-zero coefficient sampled from the sparse vector of the alleged N-point FFT.

The Matrix A in this case is adaptive to the specific signal content. In this case, the recovery process reduces to extracting the location of the non-zero (index) elements in the matrix A and use them to order the sparse K signal, embed zeros in the other locations and perform inverse FFT.

Consequently, the recovery process of the family of Sparse FFT based compression is considerably simpler than that of the compressed sensing case.

Another obvious advantage is that by deciding on the sparse coefficients beforehand, the number of sampled and transmitted frequency domain signal is optimized to either produce the lowest number of coefficients possible or to adapt the quality of the reconstructed signal (if the original signal is not perfectly sparse in the frequency domain, i.e. contains non-trivial frequency coefficients larger than K). On the other hand, the compressed sensing has to sample larger set of coefficients.

On the other hand, the SFFT requires the knowledge of the indices of the sparse frequency bins in order to reconstruct the signal. This is contrary to the CS framework that utilizes a universal sensing matrix known at both the encoder and decoder sides. In the following section we propose an innovative way to embed the indices in the extracted largest frequency bins to relax the need for extra coded values.

IV. AUDIO COMPRESSION METHODOLOGY

In this section we present the audio compression framework for both the compressive sensing based approach and the proposed application of SFFT.

The compressed sensing compression of audio signals and its respective reconstruction scheme is shown in Fig.2. First, the audio signal is divided into sequence of fixed length frames in the time domain. Each frame is transformed to the frequency domain using one of three tested transform domains, namely, wavelet transform, discrete cosine transform (DCT) and FFT. The step following the transform is applying the sensing (chosen to be random Gaussian matrix) to compress each frame by the pre-assigned ratio.

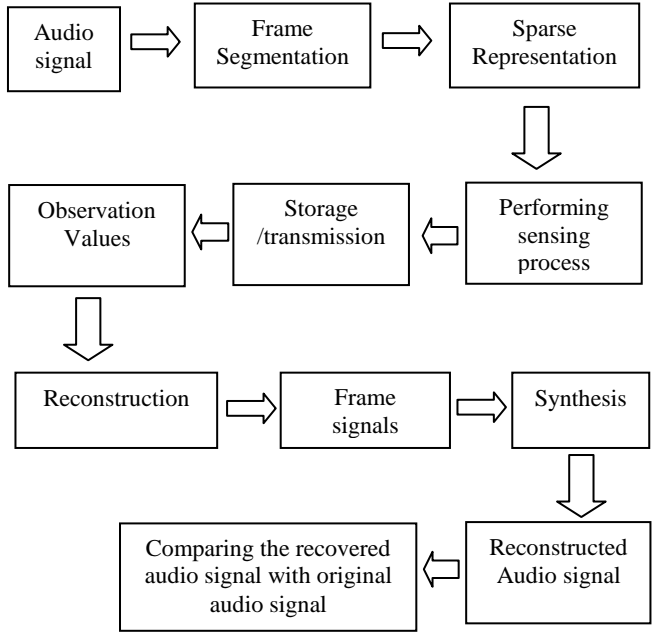

Figure 2. Audio signals compressed and reconstruction based on compressed sensing

At the decoder side, the recovery of the original signal from the sparse coefficients is done using either the OMP algorithm or the convex relaxation package. Finally, the recovered frequency domain signal is transformed back to the original time-domain representation using the adequate inverse transform.

Similar approach is proposed for the Sparse Fast Fourier Transform (SFFT) approach. The main difference, however, is that the transform applied to the signal is the FFT transform only and the choice of the sparse coefficients is done based on the intrinsic properties of the audio signal through locating the samples with the highest power content and their indices. It is worth noting that both the sparse samples and the coefficients need to be transmitted to the decoder side in order to recover the original signal. Steps as follow.

1- Dividing the audio signals into frames.
2- Apply the SFFT algorithm to extract K-largest coefficients of the signal in the frequency domain.
3- At the receiver side, using the location of K-largest coefficient and their coefficient values we reconstruct the received frames by placing the K-largest coefficient in the specified location and the rest values of the frame sets to zero.
4- Performing inverse fast Fourier transform to find time domain for each frame, then collecting these frame to reconstruct the audio signal.
5- Comparing the recovered audio signal with the original audio signal to determine the percentage of the error.

SFFT chooses large k- coefficients from the transformed signal and estimate their values and locations (indices). Then we transmit the coefficient values and their locations. This will needs more bits to represent the transmitted signal. To solve this problem we embedded the indices bits into the fractional part of transmitted coefficients. The fraction part and index is represented by n-bits. We choose only half of bits to represent the fractional part and we embedded the index bits into the least significant bits of fractional part bits (instead of representation of the fraction part by n bits, it is represented by (n/2) bit and the rest bits are the index bits). At the receiver we collect the locations of non-zero coefficients again. Transmitting the locations of non-zero coefficients will produce enhancement in the quality of reconstructed signal due to the knowledge of their locations. SFFT receiver is simpler than compressed sensing because it just needs the location of largest coefficients to construct the original signal and doing the inverse FFT on the transmitted coefficients.

V. SIMULATION RESULTS AND ANALYSIS

In this section we present the use of discrete wavelet transform (DWT), DCT and DFT in order to obtain sparse representation of our audio signal in frequency domain .also this section presents the use of SFFT algorithm to compress audio signal.

The test audio signal is a 15 second music piece and sampled at rate 44100 samples/s [20], the resulting samples are divides into frames and each frames contain 1024 samples. By applying Wavelet, DCT and DFT we obtain a sparse representation of music signal frames. Fig.3 shows the time domain and the frequency domain for the original signal.

The performance of CS and SFFT are evaluated by calculating the similarity measures [21].

$$Similarity = e^2 / E^2 \qquad (6)$$

Where $e^2$ is the matrix error given by $\|\mathbf{x} - \hat{\mathbf{x}}\|_2$, $E^2$ is $\|\mathbf{x}\|_2$. $\mathbf{x}$ is the original signal and $\hat{\mathbf{x}}$ is recovered signal.

Table 1. shows different similarity values for CS scheme and SFFT approach for different compression ratio C/R. also reconstruction time are shown in table 2.the simulations are running on Dell work-station with a processor core i7 @ 2.8 GHZ.

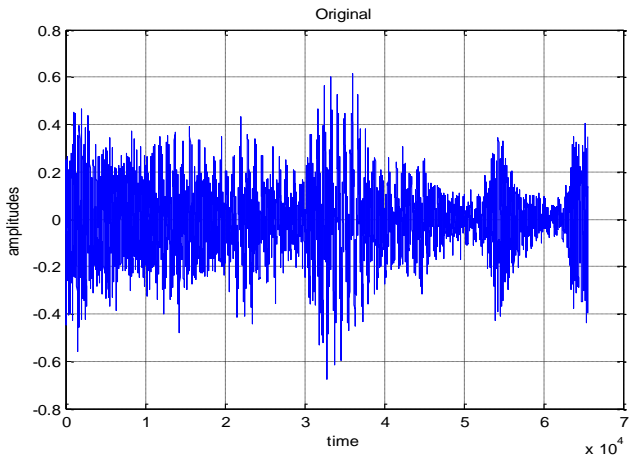

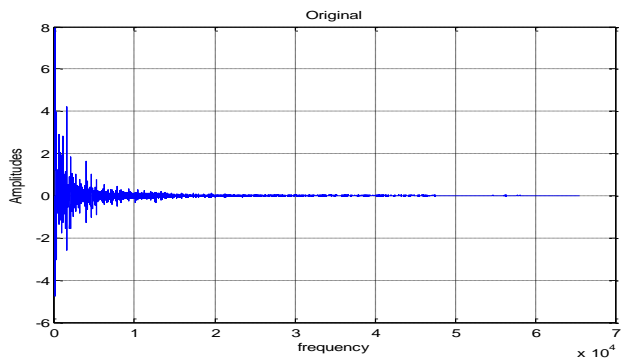

Figure 3. Audio signals (a) Time domain (b) Frequency Domain

TABLE 1. SIMILARITY VALUES BETWEEN ORIGINAL AND RECOVERED AUDIO SIGNAL

| C/R | Recovery algorithm | CS | | | SFFT Embedded index |
|---|---|---|---|---|---|
| | | DWT | DCT | DFT | |
| 20 % | OMP | 0.448 | 0.2461 | 0.0798 | 0.00009 |
| | CVX | 0.019 | 0.0033 | 0.0197 | |
| 50 % | OMP | 0.537 | 0.292 | 0.2082 | 0.00063 |
| | CVX | 0.1125 | 0.0418 | 0.0843 | |
| 60 % | OMP | 0.6068 | 0.3953 | 0.3795 | 0.0013 |
| | CVX | 0.2325 | 0.1281 | 0.1634 | |
| 66 % | OMP | .6966 | 0.5324 | 0.5577 | 0.0017 |
| | CVX | 0.3202 | 0.1916 | 0.2391 | |
| 75 % | OMP | 0.8641 | 0.7101 | 0.7310 | 0.0034 |
| | CVX | 0.5683 | 0.3709 | 0.4156 | |
| 80 % | OMP | 0.9248 | 0.8542 | 0.8678 | 0.005 |
| | CVX | 0.7471 | 0.6676 | 0.6955 | |
| 83 % | OMP | 1.0128 | 0.8934 | 0.9083 | 0.0078 |
| | CVX | 0.8534 | 0.6828 | 0.7287 | |

TABLE 2. RECONSTRUCTION TIME FOR CS AND SFFT IN SECONDS

| C/R | Recovery algorithm | CS | | | SFFT Embedded index |
|---|---|---|---|---|---|
| | | DWT | DCT | DFT | |
| 20 % | OMP | 0.0084 | 0.0084 | 0.0373 | 0.00028 |
| | CVX | 5.2525 | 5.3539 | 5.3857 | |
| 50 % | OMP | 0.0091 | 0.0088 | 0.0298 | 0.000261 |
| | CVX | 5.547 | 5.5919 | 5.6167 | |
| 60 % | OMP | 0.0084 | 0.0084 | 0.0209 | 0.00016 |
| | CVX | 5.2963 | 5.33393 | 5.3171 | |
| 66 % | OMP | 0.0096 | 0.0088 | 0.016 | 0.00017 |
| | CVX | 5.4909 | 5.5039 | 5.4685 | |
| 75 % | OMP | 0.0083 | 0.0083 | 0.0151 | 0.000153 |
| | CVX | 5.3673 | 5.3699 | 5.3337 | |
| 80 % | OMP | 0.009 | 0.0084 | 0.01 | 0.000124 |
| | CVX | 5.4727 | 5.4627 | 5.4446 | |
| 83 % | OMP | 0.0082 | 0.0084 | 0.0096 | 0.000136 |
| | CVX | 5.4911 | 5.4863 | 5.4752 | |

From table 1 shows that SFFT has better performance than CS .SFFT has lower similarity values which means lower error values between the original signal and the recovered signal comparing with CS at different compression ratio. Table 2 show that SFFT has lower reconstruction time required to recover the original signal compares to CS for all compression ratios.

The second test is providing the same number of bits at the coder output and evaluating the performance of CS and SFFT approach .in this test we choose three different values (3264, 5568, 8192 )bits. Table 3 and 4 show the similarity values and reconstruction time respectively. It is obvious from this table the performance of SFFT approach has better performance in both signal quality and reconstruction time.

TABLE 3. SIMILARITY VALUES BETWEEN ORIGINAL AND RECOVERED AUDIO SIGNAL FOR DIFFERENT NUMBER OF BITS

| C/R | Recovery algorithm | CS | | | SFFT Embedded index |
|---|---|---|---|---|---|
| | | DWT | DCT | DFT | |
| 3264 | OMP | 0.9248 | 0.8542 | 0.8678 | 0.1999 |
| | CVX | 0.7471 | 0.6676 | 0.6955 | |
| 5568 | OMP | 0.6966 | 0.5324 | 0.5577 | 0.0136 |
| | CVX | 0.3202 | 0.1916 | 0.2391 | |
| 8192 | OMP | 0.537 | 0.292 | 0.2082 | 0.0068 |
| | CVX | 0.1125 | 0.0418 | 0.0843 | |

TABLE 4. RECONSTRUCTION TIME FOR CS AND SFFT

| C/R | Recovery algorithm | CS | | | SFFT Embedded index |
|---|---|---|---|---|---|
| | | DWT | DCT | DFT | |
| 3264 | OMP | 0.009 | 0.0084 | 0.01 | 0.00013 |
| | CVX | 5.4727 | 5.4627 | 5.4446 | |
| 5568 | OMP | 0.0096 | 0.0088 | 0.016 | 0.00049 |
| | CVX | 5.4909 | 5.5039 | 5.4685 | |
| 8192 | OMP | 0.0091 | 0.0088 | 0.0298 | 0.00025 |
| | CVX | 5.547 | 5.5919 | 5.6167 | |

The results above indicate the performance gain achieved in signal reconstruction by taking the signal content into consideration. We can generalize the above results to other types of signals. Among sparse signal processing techniques, the compressed sensing provides a valuable option for sparse signal acquisition especially if low complexity is required at the acquisition. An example of the applications where the universality of compressed sensing is sensor nodes acquiring and encoding sparse data. And another application, the compressed sensing can be used in spectrum sensing to reduce the complexity of sampling process of the spectrum. In other situation Sparse FFT might be the more appealing choice. For example, if the signal is to be processed in the FFT domain and accurate estimation of the K largest coefficients is necessary, Sparse FFT should be manipulated for these types of applications. Another venture of application is when the reconstruction of FFT sparse signal is required to be lower in complexity or the bit rate is required to be optimized, while the encoder side can afford more complexity. Applications targeting better quality of the reconstructed FFT sparse signals are also another venture of application for the SFFT algorithm.

## VI. CONCLUSION

In this paper we present an index embedded SFFT-based framework to compress audio signal and comparing the results with compressed sensing results to assess the suitability of SFFT to compress audio signal. Simulation results show that SFFT-based framework outperforms the CS-based approach in signal quality. From the simulation results SFFT based approach has lower reconstruction time than CS based approach which makes it suitable for real time applications requiring lower complexity at the decoder side. The performance gain of the proposed framework justifies the need for new class of sensing matrices for the compressed sensing taking into account the intrinsic properties of the sparse signal like the perceptual effect of the signal content and a better prediction of the locations of zeros in the sparse signal.

ACKNOWLEDGMENT

This work has been supported by the Egyptian Mission of Higher Education (MoHE). I am grateful to Egypt-Japan University of Science and Technology (E-JUST) for offering the tools and equipment needed.

REFERENCES

[1] Salami, R.; Laflamme, C.; Adoul, J-p; Kataoka, A.; Hayashi, S.; Moriya, T.; Lamblin, C.; Massaloux, D.; Proust, S.; Kroon, P.; Shoham, Y., "Design and description of CS-ACELP: a toll quality 8 kb/s speech coder," *Speech and Audio Processing, IEEE Transactions on*, vol.6, no.2, pp.116,130, Mar 1998

[2] MPEG-4, ISO/IEC 14496-3:2005/Amd 2: 2006, Audio LosslessCoding (ALS), New Audio Profiles and BSAC Extensions ISO/IECStd., 2006 [Online]. Available: http://www.nue.tu-berlin.de/mp4als,2009, reference software version RM22r2

[3] Huang, H.; Rahardja, S.; Lin, X.; Yu, R.; Franti, P., "Cascaded RLS-LMS Prediction in MPEG-4 Lossless Audio Coding," *Acoustics, Speech and Signal Processing, 2006. ICASSP 2006 Proceedings. 2006 IEEE International Conference* on , vol.5, no., pp.V,V, 14-19 May 2006

[4] Ghido, F.; Tabus, I., "Sparse Modeling for Lossless Audio Compression," Audio, Speech, and Language Processing, *IEEE Transactions on* , vol.21, no.1, pp.14,28, Jan. 2013 doi: 10.1109/TASL.2012.2211014

[5] Candes, E.J.; Romberg, J.; Tao, T., "Robust uncertainty principles: exact signal reconstruction from highly incomplete frequency information," Information Theory, *IEEE Transactions on* , vol.52, no.2, pp.489,509, Feb. 2006.

[6] D. Donoho, "Compressed sensing," *IEEE Trans. Inform. Theory*, vol. 52, no. 4, pp. 1289–1306, April 2006.

[7] Michael Lustig, David Donoho, John M. Pauly, "Sparse MRI: The Application of Compressed Sensing for Rapid MR Imaging, " *Engineering in Medicine and Biology Society,EMBC, 2011 Annual International Conference of the IEEE* , vol., no., pp.5734,5737, Aug. 30 2011-Sept. 3,2011

[8] Sreenivas, T.V.; Kleijn, W.B., "Compressive sensing for sparsely excited speech signals," *Acoustics, Speech and Signal Processing, 2009. ICASSP 2009. IEEE International Conference on* , vol., no., pp.4125,4128, 19-24 April 2009.

[9] D. Giacobello, M. G. Christensen, J. Dahl, S. H. Jensen,and M. Moonen, "Sparse linear predictors for speechprocessing," *in Proc. Interspeech, 2008.*

[10] H. Hassanieh, P. Indyk, D. Katabi, and E. Price. "Simple and practical algorithm for sparse fourier transform". *ACM-SIAM Symposium on Discrete Algorithms (SODA12).*

[11] Candes, E.J.; Wakin, M.B., "An Introduction To Compressive Sampling," *Signal Processing Magazine, IEEE* , vol.25, no.2, pp.21,30, March 2008.

[12] Shah, P.; Grant, S.L.; Benesty, J., "On an iterative method for basis pursuit with application to echo cancellation with sparse impulse responses," *Acoustics, Speech and Signal Processing (ICASSP), 2012 IEEE International Conference on* , vol., no., pp.177,180, 25-30 March 2012.

[13] Michael Grant and Stephen Boyd. CVX: Matlab software for disciplined convex programming, version 2.0 beta.http://cvxr.com/cvx, September 2012.

[14] Michael Grant and Stephen Boyd. Graph implementations for nonsmooth convex programs, Recent Advances in Learning and Control (a tribute to M. Vidyasagar), V. Blondel, S. Boyd, and H. Kimura, editors, pages 95-110, Lecture Notes in Control and Information Sciences,Springer,2008. http://stanford.edu/~boyd/graph_dcp.html.

[15] Tropp, J.A.; Gilbert, A.C., "Signal Recovery From Random Measurements Via Orthogonal Matching Pursuit," ,*IEEE Trans Information Theory* , vol.53, no.12, pp.4655,4666, Dec. 2007.

[16] Donoho, D.L.; Tsaig, Y.; Drori, I.; Starck, J-L, "Sparse Solution of Underdetermined Systems of Linear Equations by Stagewise Orthogonal Matching Pursuit," Information Theory, *IEEE Transactions on* , vol.58, no.2, pp.1094,1121, Feb. 2012.

[17] Deanna Needell and Joel A. Tropp, "Cosamp: iterative signal recovery from incomplete and inaccurate samples," *Commun. ACM*, vol. 53, no. 12, pp. 93-100, Dec. 2010.

[18] Robert Tibshirani, "Regression shrinkage and selection via the lasso," *Journal of the Royal Statistical Society. Series B (Methodological)*, vol. 58, no. 1, pp. 267 -288, 1996. .

[19] A. Gilbert, M. Strauss, J. Tropp, and R. Vershynin. One sketch for all: Fast algorithms for compressed sensing. *In Proc. 39th ACM Symp. Theory of ComputingSan Diego*, June 2007.

[20] http://cse.unl.edu/~sincovec/Matlab/Assignments/Project/Sound_Editor_Project/Sound_Editor_Implementation.htm

[21] R.G. Moreno-Alvarado, Mauricio Martinez-Garcia, "DCT-compressive Sampling of Frequency-sparse Audio Signals,"*Proceedings of the World Congress on Engineering 2011* Vol II WCE 2011, July 6 - 8, 2011, London, U.K